\documentclass{emulateapj}
\usepackage{apjfonts}
\usepackage{lscape}

\newcommand{\gsim}{>}

\newcommand{\dfn}{D$_{n}(4000)$}
\newcommand{\ha}{H$\alpha~$}

\newcommand{\hda}{H$\delta_{\mbox{A}}$}

\newcommand{\airx}{$\widehat{A}_{IRX}$}

\newcommand{\msun}{M$_\sun$}

\begin{document}
\title{Ultraviolet through Infrared Spectral Energy Distributions from 1000 SDSS Galaxies: Dust Attenuation}

\author{
Benjamin D. Johnson\altaffilmark{1}
David Schiminovich\altaffilmark{1},
Mark Seibert\altaffilmark{2},
Marie Treyer\altaffilmark{3,4},
D. Christopher Martin\altaffilmark{4},
Tom A. Barlow\altaffilmark{4}, 
Karl Forster\altaffilmark{4},
Peter G. Friedman\altaffilmark{4},
Patrick Morrissey\altaffilmark{4},
Susan G. Neff\altaffilmark{5},
Todd Small\altaffilmark{4},
Ted K. Wyder\altaffilmark{4},
Luciana Bianchi\altaffilmark{6},
Jose Donas\altaffilmark{3},
Timothy M. Heckman\altaffilmark{7},
Young-Wook Lee\altaffilmark{8},
Barry F. Madore\altaffilmark{2},
Bruno Milliard\altaffilmark{3},
R. Michael Rich\altaffilmark{9},
Alex S. Szalay\altaffilmark{7},
Barry Y. Welsh\altaffilmark{10}, 
Sukyoung K. Yi \altaffilmark{8}
}

\altaffiltext{1}{Department of Astronomy, Columbia University, New York, NY 10027}
\altaffiltext{2}{Observatories of the Carnegie Institution of Washington, 813 Santa Barbara St., Pasadena, CA 91101}
\altaffiltext{3}{Laboratoire d'Astrophysique de Marseille, BP 8, Traverse du Siphon, 13376 Marseille Cedex 12, France}
\altaffiltext{4}{California Institute of Technology, MC 405-47, 1200 East California Boulevard, Pasadena, CA 91125}
\altaffiltext{5}{Laboratory for Astronomy and Solar Physics, NASA Goddard Space Flight Center, Greenbelt, MD 20771}
\altaffiltext{6}{Center for Astrophysical Sciences, The Johns Hopkins University, 3400 N. Charles St., Baltimore, MD 21218}
\altaffiltext{7}{Department of Physics and Astronomy, The Johns Hopkins University, Homewood Campus, Baltimore, MD 21218}
\altaffiltext{8}{Center for Space Astrophysics, Yonsei University, Seoul 120-749, Korea}
\altaffiltext{9}{Department of Physics and Astronomy, University of California, Los Angeles, CA 90095}
\altaffiltext{10}{Space Sciences Laboratory, University of California at Berkeley, 601 Campbell Hall, Berkeley, CA 94720}

\begin{abstract}
The meaningful comparison of models of galaxy evolution to observations is critically dependent on the accurate treatment of dust attenuation. To investigate dust absorption and emission in galaxies we have assembled a sample of $\sim$1000 galaxies with ultraviolet (UV) through infrared (IR) photometry from \emph{GALEX}, SDSS, and \emph{Spitzer} and optical spectroscopy from SDSS. The ratio of IR to UV emission (IRX) is used to constrain the dust attenuation in galaxies.  We use the 4000\AA~break as a robust and useful, although coarse, indicator of star formation history (SFH).   We examine the relationship between IRX and the UV spectral slope (a common attenuation indicator at high-redshift) and find little dependence of the scatter on \dfn.  We construct average UV through far-IR spectral energy distributions (SEDs) for different ranges of IRX, \dfn, and stellar mass (M$_{*}$) to show the variation of the entire SED with these parameters.  When binned simultaneously by IRX, \dfn, and M$_{*}$ these SEDs allow us to determine a low resolution average attenuation curve for different ranges of M$_{*}$.  The attenuation curves thus derived are consistent with a $\lambda^{-0.7}$ attenuation law, and we find no significant variations with M$_{*}$.  Finally, we show the relationship between IRX and the global stellar mass surface density and gas-phase-metallicity. Among star forming galaxies we find a strong correlation between IRX and stellar mass surface density, even at constant metallicity, a result that is closely linked to the well-known correlation between IRX and star-formation rate.

\end{abstract}

\keywords{galaxies:fundamental parameters --- galaxies:evolution --- dust:extinction --- ultraviolet:galaxies --- infrared:galaxies}

\section{Introduction}
Measurement of the star formation history (SFH) of galaxies, and the distribution thereof, tests models of galaxy evolution.  Such measurement requires the detailed treatment of dust attenuation, which poses one of the most severe obstacles to converting observed properties (e.g. broadband colors or line fluxes) into a SFH or star formation rate (SFR). 

One of the primary tools for the measurement of dust attenuation, especially at high-redshift, is the relation between the ratio of infrared flux to ultraviolet flux (IRX) and the ultraviolet spectral slope ($\beta$, where $f_{\lambda}\sim\lambda^{\beta}$).  The former can be considered an approximate measure of the amount of dust attenuation since the IR flux is produced by light absorbed at primarily UV wavelengths, while the UV flux measures the light transmitted at these wavelengths.  The UV spectral slope measures the reddening (or color excess) of the UV spectrum due to selective absorption by dust -- assuming that before attenuation the spectrum is nearly flat and relatively insensitive to the SFH.  The utility of this relation lies in the relatively easy measurement of the restframe UV color for large samples of galaxies at $z>1$, from which the attenuation can be inferred.

The calibration (and scatter) of this relation is of great interest for measurements of the SFR of galaxies at high redshift (especially those based on the restframe UV luminosity) and for the resulting estimates of the SFR density and its evolution \citep{madau96, ds05}.  This calibration is usually empirical, as is necessary at least in part because of the dependence of the reddening effect of the dust on the assumed properties of the dust (e.g., the extinction law and dust geometry).  One must then assume that these dust properties do not evolve strongly with redshift, although calibrations based entirely on high-redshift galaxies may soon become available \citep{reddy_dust}. 

Such empirical calibration, at any redshift, requires both restframe IR and UV observations galaxies.  In the past, such data has been available from \emph{IRAS} and \emph{IUE} for samples of nearby starbursting galaxies, where a strong correlation between IRX and $\beta$ was found \citep{MHC}.  With the launch of \emph{GALEX}, UV data has become available for a larger sample of more normal galaxies \citep{seibert05a, buat05}, for which it appears that the relation between IRX and $\beta$ is shifted to redder colors, and has increased scatter.  Similar results were obtained for a sample of individual OB associations by \citet{bell02b}, who suggested that these effects were due to variations in the relative star-dust geometry \citep[e.g.,][]{WG00}.  \citet{kong04} found a trend with the 4000\AA~break (a coarse measure of the SFH) of the offset of starburst galaxies from the best fit IRX-$\beta$ relation. On the basis of spectral synthesis modelling they thus ascribed the scatter and shifts in towards redder colors in the IRX-$\beta$ relation to the effect of SFH on $\beta$, in the sense that a large population of old stars will tend to produce a redder intrinsic, unattenuated, UV color.  The largest effects in the models were seen for significant old bursts superimposed on a smoother SFH.  The spatial resolution of \emph{GALEX}, combined with the spatial resolution of \emph{Spitzer}, has allowed for the investigation of the IRX-$\beta$ relation in detail in nearby, resolved galaxies \citep{boissier04, gil_de_paz, calzetti05}.  One important outcome of this work is the suggestion that populations of different ages are obscured by differing amounts of dust \citep{calzetti05}, a result consistent with the findings of \citet{charlot00} for global measures of attenuation. Such a scenario further complicates the interpretation of the IRX-$\beta$ relation \citep{panuzzo07}.  The combination of sensitive UV and IR observations of galaxies with \emph{GALEX} and \emph{Spitzer} has enabled the investigation of the IRX-$\beta$ relation for large samples of galaxies \citep{buat05, cortese06}.

This paper is part of a series investigating the UV through IR properties of a large, well-defined sample of galaxies observed spectroscopically for SDSS, and by \emph{GALEX} and \emph{Spitzer}.  The additional diagnostics of stellar populations, SFR, and attenuation provided by the SDSS data make the sample presented here a unique and important testbed for the understanding of the IRX-$\beta$ relation. 

The large amount of homogenous ancillary information available from the SDSS observations, in addition to the UV and IR data, also allows us to investigate the effects of dust across the UV through near-IR spectrum.  When such a wide range of wavelengths is considered, the effects of SFH on the spectrum become much more pronounced than they are for the UV spectrum alone \citep{paper1}.  Indeed, optical colors are often used as a proxy for SFH \citep{bell04_red, faber05}, despite the potentially large additional contribution of attenuation to these colors. Here we present the global UV through IR spectral energy distributions (SEDs) of a large range of galaxy types. We show how these measured SEDs vary as a function of stellar mass, SFH, \emph{and} attenuation.  In particular, we use these SEDs to derive an `average' dust attenuation law, and make a first attempt to determine the variation of this law (if any) with the stellar mass of the galaxy.  These SEDs are complementary to the integrated UV through IR SEDs presented for the smaller samples of much more local, resolved galaxies presented by \citet{gil_de_paz} and \citet{dalex}, and to the average UV through IR SEDs of higher redshift galaxies \citep{zheng07}.

Besides constraining the correction of derived physical values for dust attenuation, the relation of the dust attenuation to the SFH, SFR, and stellar mass may provide an additional constraint on models of galaxy evolution, as the attenuation should be proportional to the gas surface density $\Sigma_{gas}$ and gas-phase metallicity $Z$ \citep{WH96, bell03_radio, martin07, cortese06}. Models that self-consistently treat absorption and emission by dust are necessary and must be able to both predict the correct distribution of $\Sigma_{gas} Z$ \citep[as are given by the models of, e.g.][]{croton06, delucia06, somerville01} {\it and} convert this quantity into a dust attenuation curve to accurately predict the UV through IR properties of galaxies. We investigate the relation between $\Sigma_{gas} Z$ and attenuation for the galaxies in our sample that have metallicity measurements.

\section{Data}
In this study we use optical spectroscopic and photometric observations of galaxies from SDSS, UV observations of these galaxies from \emph{GALEX}, and IR observations of these galaxies from \emph{Spitzer}.  Readers are referred to \citet{paper1} for details of the data reduction and sample definition.  Here we provide a summary of the sample properties and briefly describe several parameters derived from the data that are used throughout this analysis.

The sample consists of galaxies targeted spectroscopically by SDSS (thus having a magnitude $r<17.7$), for which a stellar mass has been determined by \citet{kauffmann03a} and which have been observed by both \emph{GALEX} and \emph{Spitzer}.  This sample is located in two regions of the sky, the Lockman Hole (observed by \emph{Spitzer} as part of the SWIRE survey) and the Spitzer Extragalactic First Look Survey (FLS).  Table \ref{tbl:obs} gives the number of galaxies observed in each field and the detection rates. The sample properties (redshift, stellar mass, and morphological distribution, etc.) are very similar to the sample of \citet{kauffmann03a}, although the redshift distribution is less smooth due to the presence of groups, filaments, and sheets in the spatial distribution of galaxies in these fields. The median redshift of the sample is $z=0.11$ with a maximum of $z<0.305$. The median stellar mass (see below) is $\log\mbox{M}_{*}=10.7$\msun with a range of $8.45 < \log\mbox{M}_{*} < 11.91$.  The SFR range is $sim 0.01$-$60$\msun$\cdot yr^{-1}$ \citep{paper2, brinchmann04}. A detailed investigation of the morphological properties of the ample is beyond the scope of this work, but we note that irregular, elliptical, and spiral galaxies of various inclinations (face-on through edge-on) are included \citep[see][Fig. 4]{paper1}. The SDSS spectroscopic galaxies with stellar masses from \citet{kauffmann03a} that are \emph{not} detected in the UV or IR are $\sim30$\% of galaxies on the red sequence and a few blue, low stellar mass dwarf galaxies that are missed primarily because we require objects to have Petrosian radii in the $r$ band of less than 11\arcsec and redshift $z>0.02$, insuring accurate photometry. Photometry is from the modified SExtractor output for the \emph{GALEX} far-UV ($f$, $\lambda \sim 1528$\AA) and near-UV ($n$, $\lambda \sim 2271$\AA) data, SDSS Petrosian magnitudes ($ugriz$) for the optical data, and large aperture photometry for the \emph{Spitzer} data (with 7\arcsec radius in the IRAC 3.6\micron~through 7.8\micron~bands, 12\arcsec radius at 24\micron, and 16\arcsec radius at 70\micron~where the sources are largely unresolved).  The $f$ through 3.6\micron magnitudes are $K$-corrected to $z=0.1$ (the median redshift of the sample) via the method of \citet{blanton_k}, with the result denoted by a superscripted $0.1$, e.g. $^{0.1}u$. We also determine $K$-corrections of the $f$ and $n$ magnitudes to $z=0.0$ ($^{0.0}f$, $^{0.0}n$) for comparison to more local galaxies (\S\ref{sec:comp}).


\begin{deluxetable}{cccccc}
\tablecolumns{6}
\tablecaption{Observations \label{tbl:obs}
}
\tablehead{
\colhead{Field Name} & \colhead{Size}  & \colhead{N$_{obs}$} & \colhead{N$_{det}$} & \colhead{N$_{det}$} & \colhead{N$_{smpl}$} \\
\colhead{}   & \colhead{(deg$^{2}$)} & \colhead{} & \colhead{$f,n<25$} & \colhead{$m_{24}<19.5$} &  \colhead{}
}
\startdata
Lockman Hole & $\sim 9$ & 872 & 792 & 819 & 721  \\
FLS & $\sim 3$ & 186 & 147 & 158 & 118 
\enddata

\end{deluxetable}

In this work we are interested in the stellar masses derived by \citet{kauffmann03a} and the metallicities derived by \citet{tremonti04}. These properties have been made available as catalogs\footnote{http://www.mpa-garching.mpg.de/SDSS/} by these authors. The stellar masses are derived from the $z$-band SDSS photometry.  The mass-to-light ratio in that band is obtained for each galaxy from fits of stellar population synthesis models to the optical spectrum, in particular the 4000\AA break strength and the H$\delta$ absorption strength, \hda.  The gas phase metallicities are derived from the SDSS measured emission lines for galaxies with strong emission lines that do not show evidence of AGN activity in the optical \citep[see][for details]{tremonti04}.  We also consider in \S\ref{sec:sigma_z} the SFR derived by \citet{brinchmann04} from a comparison of measured optical emission lines to a large suite of models of galaxy spectra.

\subsection{\dfn}

The 4000\AA~break strength, here defined as in \citet{balogh99} (\dfn), is a useful indicator of the mass-to-light ratio since it is sensitive to the star-formation history (SFH). Larger \dfn~signifies, approximately, a larger ratio of old stars to young stars.  This break is measured at high signal-to-noise ratio (S/N) for all of the galaxies in the sample, and is not very sensitive to the presence of dust attenuation \citep[but see][]{macarthur05}.  \dfn~is only measured within the 3'' radius spectroscopic aperture of the SDSS, and can therefore be overestimated for galaxies with moderate bulge/disk ratios \citep{kauffmann06}.  \dfn~serves as our primary measure of SFH.


\subsection{IRX}
\label{sec:irx}
The combination of UV and IR data allows us to construct a robust measure of attenuation: the so-called infrared excess, IRX$=\log (\mbox{L}_{dust}/\mbox{L}_{uv})$. We define $\mbox{L}_{uv}=\nu L_{\nu}$ where $\nu=c/1390\mbox{\AA}$ ($\lambda=1390$\AA is the effective wavelength of the $^{0.1}f$ band) and calculate L$_{\nu}$ from the $K$-corrected absolute magnitude, assuming the redshift-distance relation given by a concordance cosmology with $\Omega_{m}=0.3, \Omega_{\Lambda}=0.7, \mbox{H}_{0}=70$ km s$^{-1}$ Mpc$^{-1}$. $\mbox{L}_{dust}$, the 8-1000\micron~luminosity, is estimated from the 24\micron~luminosity as in \citet{paper1}, with a bolometric correction that depends on the ratio of 8\micron~to 24\micron~fluxes.  These bolometric corrections are derived from the models of \citet{dale01}. 

If the IR emission traces the (predominantly blue) light from (predominantly) young stars that is absorbed by dust and reradiated, and the UV emission traces the light from young blue stars that is transmitted, then their ratio is a measure of the optical depth. The UV attenuation in magnitudes can be written in terms of IRX, including a parameter $\eta$ to account for light absorbed by dust at wavelengths other than the UV, as
\begin{equation}
\label{eqn:airx}
\widehat{A}_{IRX}=2.5\log (\eta 10^{IRX}+1) 
\end{equation}
with $\eta=1/1.68$ \citep{MHC,paper2}. Other conversions between IRX and attenuation are available \citep[e.g.,][]{bell02b, buat05} but are consistent with this formulation, which is physically motivated and simply understood.  However, this formulation relies on a number of simplifying assumptions.  First, the fraction of light absorbed by dust at wavelengths other than $\lambda\sim 1390$\AA~may depart from the canonical value of 1$-$1/1.68, depending especially on the SFH of the galaxy \citep[see, e.g.,][]{paper2} -- this will change the relation between \airx~and the true attenuation at 1390\AA, $A_{1390}$. Related to this, different stellar populations \emph{within} a galaxy may be attenuated by different amounts of dust \citep{charlot00, calzetti05}.  Second, the relative geometry of the stars and dust can lead to the misestimation of $A_{1390}$ by A$_{IRX}$, with an extreme example given by a group of stars surrounded by a ring of dust normal to the line of sight \citep{bell02a}.  This effect has been examined in detail by \citet{WG00}, \citet{buat96}, \citet{gordon00}, and \citet{pierini04}.  Nevertheless, IRX is a qualitatively different measure of attenuationfrom ones that rely on a color excess \citep[e.g. the Balmer decrement][and references therein]{kennicutt98}.

\section{Attenuation from the IRX-$\beta$ Relation}
\label{sec:irx_beta}

The relation between dust attenuation and UV color, the IRX-$\beta$ relation (where $\beta$ gives the power-law exponent of the UV spectrum, $f_{\lambda}\sim\lambda^{\beta}$) is a common tool for the analysis of the attenuation in galaxies at high-redshift (and low).  The restframe UV color is more easily measured at high-redshift, where it is redshifted into the optical, than many other diagnostics (e.g. the Balmer decrement).  However, at low redshift the calibration has been restricted, until recently, to galaxies with very high SFR \citep{MHC}. A number of more recent studies have suggested that the relation becomes significantly scattered when extended to less rapidly star-forming galaxies, or even to individual OB associations, and is shifted to redder UV color \citep{bell02a,seibert05a, cortese06, gil_de_paz, boissier06, dalex}. \citet{kong04} suggest, on the basis of stellar population synthesis modeling, that the increased scatter and the shift to redder UV color, may be due to the effect of SFH on $\beta$. \citet{bell02b} argue that these effects are due to radiative transfer through non-trivial dust geometries. \citet{panuzzo07} find that differing attenuation of stellar populations with different ages can lead to significant changes in the location of galaxies in the IRX-$\beta$ plane.


\begin{figure*}[t]
\epsscale{1.0}
\plotone{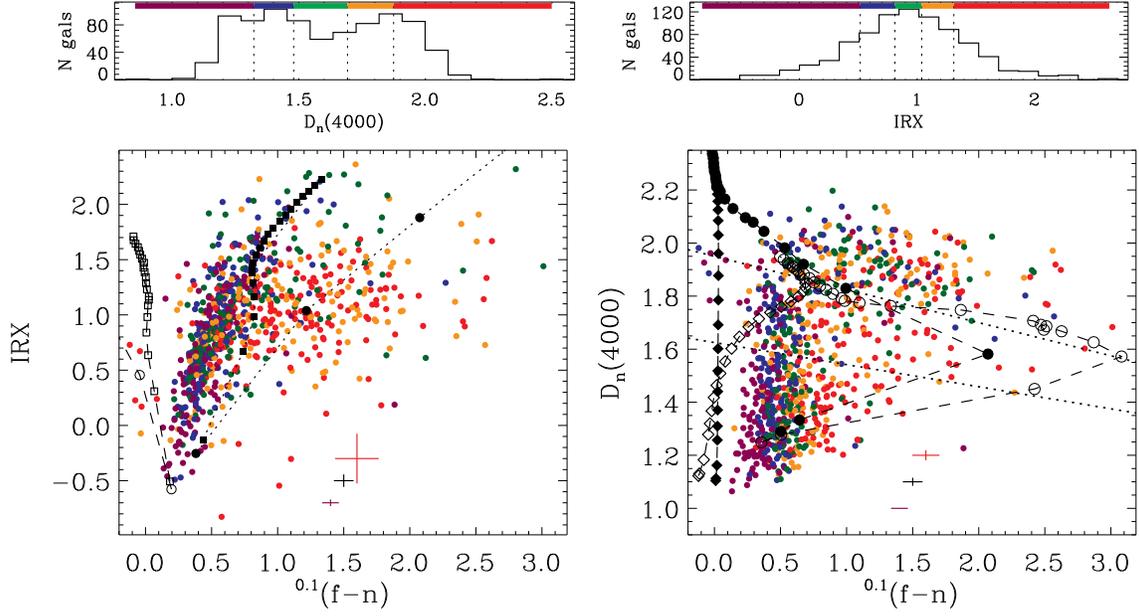}
\figcaption{The IRX-$\beta$ relation shown in two projections.  {\it Left:} IRX vs. $^{0.1}(f-n)$ color. The color of each point indicates its \dfn~quintile from lowest ($purple$) to highest ($red$). Error bars at lower left show the median errors for the entire sample ($black$) as well as the lowest and highest \dfn~quintiles ($purple$ and $red$, respectively). The black symbols and dashed lines show the effect of dust attenuation from \citet{WG00}, with a SHELL global geometry and for combinations of a MW ($open symbols$) or SMC ($filled symbols$) extinction law and a clumpy ($squares$) or homogeneous ($circles$) local dust distribution.  Symbols run from $\tau_V=0.1$ to $10$ in steps of 0.5.  {\it Right:}    \dfn~vs. $^{0.1}(f-n)$ color.  The color of each point indicates its IRX quintile form lowest (purple) to highest (red).  Overplotted in black are four \citet{BC03} population synthesis models with Z/Z$_{\sun}=0.4$ ($open$), $2.5$ ($filled$) and exponentially declining SFR with time constants $\tau_{SF}=0.1$ ($circle$), $1$ ($diamond$). The symbols are placed at intervals of 500 Myr, from 0.5 to 13 Gyr.  The dotted lines show Eqs \ref{eqn:old_cut1} and \ref{eqn:old_cut2}.
\label{fig:irx_beta}}
\end{figure*}

\subsection{As a Function of  \dfn}
\label{sec:irx_beta1}

We construct an analog of the IRX-$\beta$ relation for our sample galaxies.  We use $^{0.1}(f-n)$ color as a proxy for $\beta$, since we lack UV spectra -- see \citet{kong04} for a comparison of $\beta$ derived from \emph{GALEX} colors and from spectra. In the left panel of Figure \ref{fig:irx_beta} we show the relation between IRX and $^{0.1}(f-n)$ color for different ranges of \dfn.  We find that the overall scatter in the IRX-$\beta$ relation is much reduced compared to \citet{dissecting}.  This is due to the much smaller errors in UV color with the $\gsim 10$ times deeper UV data presented here.  However, there is still no clear trend of the scatter in the IRX-$\beta$ relation with \dfn~(our chosen SFH indicator) for \dfn~$< 1.7$.  Such an effect is constrained by these data to be quite small for low \dfn, although it may also be masked by a strong dependence of the conversion between IRX and the true attenuation on \dfn~\citep{paper2}.

As seen in \citet{dissecting}, for \dfn$> 1.7$ the scatter in the relation is greatly increased, and the galaxies have in general much redder UV color than they do for \dfn~$< 1.7$. \citet{gil_de_paz} see a similar scatter in very UV red galaxies. There is no strong dependence of the behaviour at \dfn$>1.7$ or \dfn$<1.7$ on the SFH indicator being used (e.g. the \ha equivalent width, see \citet{paper2} for examples of additional SFH indicators available for this sample). 

We also show in Figure \ref{fig:irx_beta} the expectation for the relation between \airx~and $^{0.1}(f-n)$ color for a variety of dust models drawn from \citet{WG00}.  See \citet{paper1} for a detailed description of these models and their implementation in these color-color diagrams.  Note that no single model matches the data clearly - in particular models including a Milky Way like extinction curve are strongly disfavored. The 2175\AA~bump in such extinction laws, thought to be due to absorption by ploycyclic aromatic hydrocarbon (PAH) molecules \citep{draine_li06}, leads to very little reddening of the UV color with IRX, and perhaps even \emph{bluer} colors with increasing attenuation. Note that small variations between galaxies of the particular star/dust geometry, as well as the dust \emph{extinction} law, can easily lead to significant scatter in the IRX-$\beta$ diagram.  One must also consider here the possible effects of a dust attenuation optical depth that varies within the galaxy as a function of the stellar population \citep{charlot00}, which might include minimizing the effect of the 2175\AA~bump on the color excess as well as significant offsets in the IRX-$\beta$ plane \citep{panuzzo07}.  Finally, despite the deeper UV data there are still considerable measurement errors in the UV color, and also measurement and systematic errors affecting the calculated IR luminosity \citep{paper1}, that can easily lead to considerable scatter. 

The right panel of Figure \ref{fig:irx_beta} shows that for a smoothly declining exponential star formation model (SFR$\sim e^{-t/\tau_{SF}}$) with time constant $\tau_{SF}=1$ Gyr \citep{BC03, paper1}, the UV color does not appear to change significantly even as \dfn~increases, except for low-metallicity models at late times and high \dfn.  A correlation between \dfn~and unattenuated UV color at early times would seem to require the presence of significant, rapidly declining bursts (e.g. the $\tau_{SF}=0.1$ Gyr model in Fig. \ref{fig:irx_beta}).  Indeed, the models of \citet{kong04} that show the largest deviations from the starburst relation between IRX and $\beta$ are those containing significant, short bursts of star formation -- such SFHs may not be well-represented in this sample.

The right panel of Figure \ref{fig:irx_beta} also shows that the UV colors of many of the galaxies with red $^{0.1}(f-n)$ color and large \dfn~are consistent with being produced entirely by evolved stars from old, passively evolving systems \citep[see also][]{donas_red}.  As we show in \citet{paper2} there may also be large contributions to the IR luminosity from dust heated by old, redder stars.  Thus, the meaning of IRX in these galaxies is unclear.  For comparison with other work we define two cuts, meant to remove such galaxies for which IRX is likely be a poor measure of attenuation because of the contribution of old stars to both the UV and IR flux.  These are
\begin{eqnarray}
\label{eqn:old_cut1}
\mbox{D}_{\mbox{n}}(4000) & < & 1.95-0.125\times ^{0.1}(f-n)\\
\label{eqn:old_cut2}
\mbox{D}_{\mbox{n}}(4000) & < & 1.625-0.833\times ^{0.1}(f-n)
\end{eqnarray}

\begin{figure}[t]
\epsscale{1.0}
\plotone{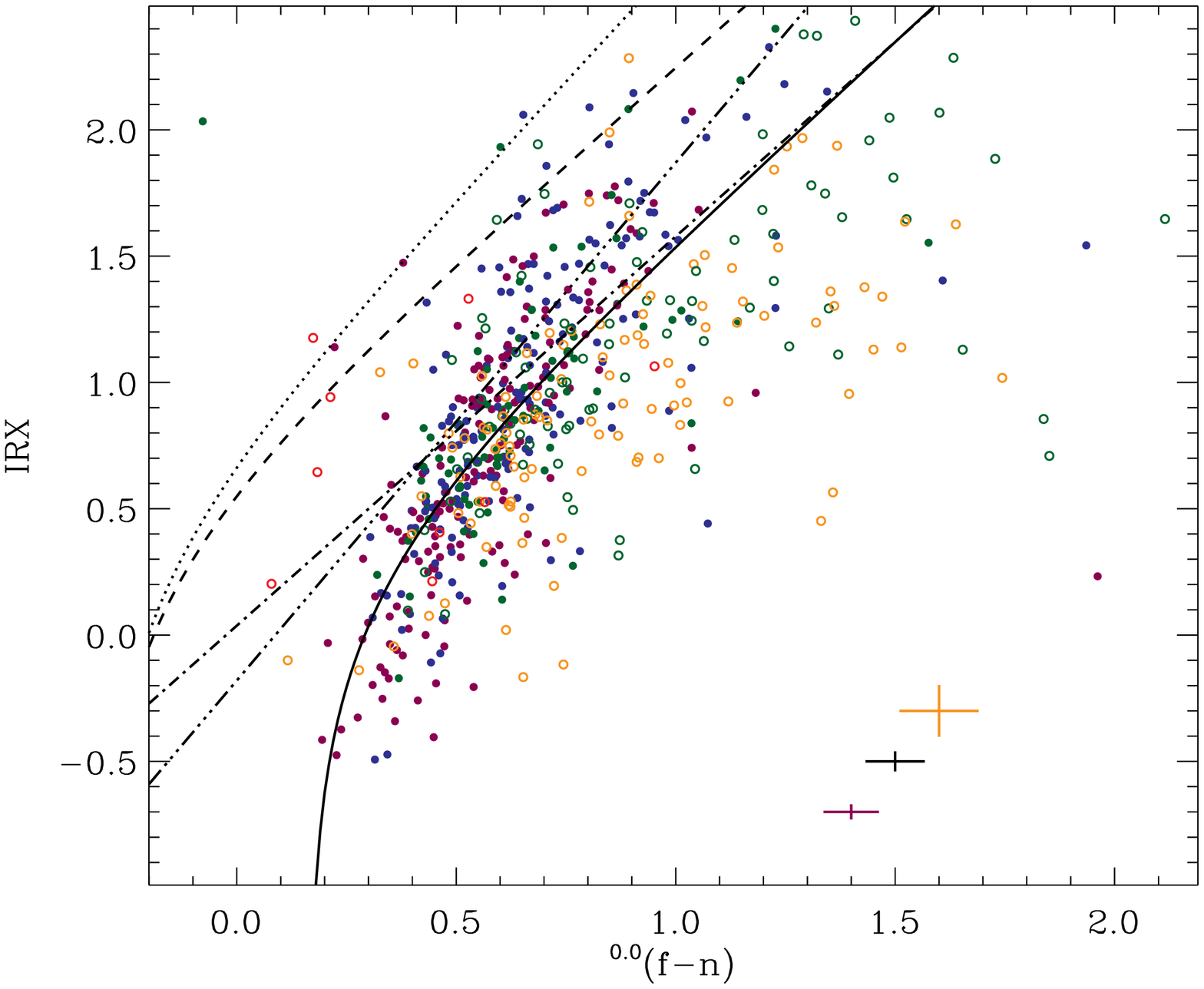}
\figcaption{IRX vs. $^{0.0}(f-n)$ color (an analog of the IRX-$\beta$ relation).  The $K$-corrections to $z=0$ are determined following \citet{blanton_k}.  The IRX-$\beta$ relations of \citet{kong04} ($dotted line$) ,\citet{seibert05a} (shifted redward by 0.2 mag; $solid line$), \citet{MHC} ($dashed line$), \citet{cortese06} (Eq \ref{eqn:cortese}; $dot-dashed line$), and \citet{gil_de_paz} (Eq \ref{eqn:gil_de_paz}; $triple-dot-dashed line$) are overplotted  (see text for details).  Only those galaxies that satisfy eq \ref{eqn:old_cut1} are shown. Galaxies that do not satisfy eq \ref{eqn:old_cut2} are shown as open circles. The colors of the points are as in Fig. \ref{fig:irx_beta}, with \dfn~quintiles determined from the entire sample (see Figure \ref{fig:irx_beta}). Error bars in the lower right show the median error for the lowest quintile of \dfn~(purple) and second highest quintile (orange), as well as the median error for all galaxies shown (black).  
\label{fig:irx_beta_z0}}
\end{figure}

\subsection{Comparison with Other Studies}
\label{sec:comp}
A comparison with the IRX-$\beta$ relations of \citet{MHC}, \citet{kong04}, \citet{seibert05a}, \citet{cortese06}, and \citet{gil_de_paz} is possible using the conversions from $^{0.0}f-n$ color to $\beta$ and from IRX to $A_{fuv}$ given in \citet{kong04} and \citet{seibert05a}.  We collect the various relations here for clarity:
\begin{eqnarray}
\label{eqn:mhc}
A_{fuv} & = & 4.37+1.74\beta_{IUE}\\
\label{eqn:mark}
A_{fuv} & = & 3.79+1.74\beta_{sb}\\
\label{eqn:kong}
A_{fuv} & = & 5.25+2.125\beta_{k}
\end{eqnarray}


\begin{figure}
\epsscale{1.0}
\plotone{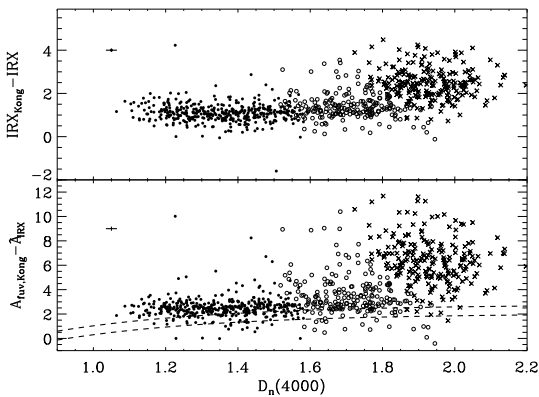}
\figcaption{{\it Top}: Deviation in IRX from that predicted by the relation of \citet{kong04}, given the measured $^{0.0}(f-n)$ color, shown as a function of \dfn.  Filled circles are galaxies that satisfy eq. \ref{eqn:old_cut2}, open circles are galaxies that satisfy eq. \ref{eqn:old_cut1} but not eq. \ref{eqn:old_cut2}, and crosses are galaxies that do not satisfy eq. \ref{eqn:old_cut1}.  {\it Bottom}: As in the top panel, but expressed in terms of magnitudes of attenuation.  The dashed lines show the predicted offsets as a function of \dfn~from \citet{kong04} (using the the approximations given in eqs. 8 and 9 of that work) for $^{0.0}(f-n)=0.2$ ($bottom curve$) and $^{0.0}(f-n)=1.5$ ($top curve$).  Error bars at top left show the median errors for the sample.
\label{fig:irx_beta_delts}}
\end{figure}

\noindent where $\beta_{IUE}$ is measured from \emph{IUE} spectra, $\beta_{sb}=2.286(f-n)-2.096$ \citep{seibert05a}, $\beta_{k}=2.201 (f-n)-1.804$ \citep{kong04}, we assume $\beta_{IUE}=\beta_{k}$, and we use the conversion from $A_{fuv}$ to IRX given in equation \ref{eqn:airx}, as was done by these authors.  The \citet{kong04} and \citet{MHC} relations are based on a sample of starbursting galaxies observed with \emph{IUE} and \emph{IRAS}.  The \citet{seibert05a} relation is derived from a sample of galaxies of a wider range of types, observed with \emph{GALEX} and \emph{IRAS}.  An important difference between this study and that of \citet{seibert05a} is that the UV color calibration of \emph{GALEX} has changed by $\sim0.2$ mag such that galaxies now have larger $f-n$, (\citet{galex_pipe}, M. Seibert, 2007, private communication).  We thus show the relation of \citet{seibert05a} shifted by 0.2 mag to the red; there is a small effect ($<0.1$) of these different calibrations on IRX, which we ignore.  In addition to the three relations given above, \citet{cortese06} and \citet{gil_de_paz} give relations that are linear between IRX and $\beta$ or $(f-n)$, derived from a sample of nearby galaxies with a wide range of SF properties.  These are:
\begin{eqnarray}
\label{eqn:cortese}
\mbox{IRX} & = & 1.30+0.7\beta_{k} \\
\label{eqn:gil_de_paz}
\mbox{IRX} & = & -0.18+2.05 (f-n) 
\end{eqnarray}


\begin{figure*}[t]
\epsscale{1.0}
\plotone{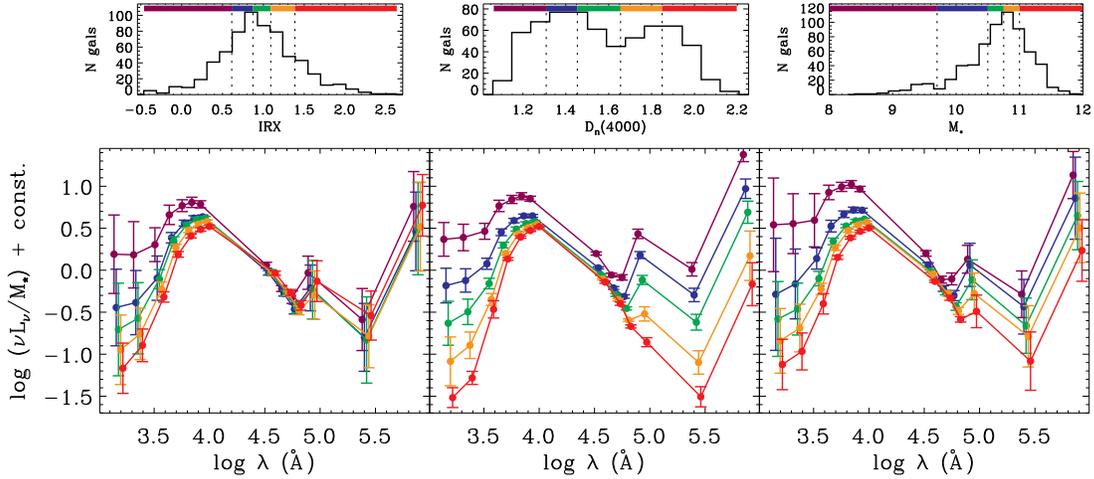}
\figcaption{SEDs normalized by stellar mass and averaged in bins of several galaxy parameters.  From left to right these parameters are IRX, \dfn, and M$_{*}$.  The bins used for color coding are shown above each panel.  Error bars give the dispersion at each wavelength band of $\log M_{*}/\nu L_{\nu}$ within each parameter bin -- the distribution of $M_{*}/\nu L_{\nu}$ is nearly log-normal for the parameter bins chosen here.  For $\nu L_{\nu}$ and M$_*$ in solar units the constant on the $y$-axis is 0.99, as for the values in Table \ref{table:sed}.
\label{fig:sed}}
\end{figure*}

These five relations are shown in Figure \ref{fig:irx_beta_z0}.  There is reasonable agreement with the (shifted) relation of \citet{seibert05a}, especially at $^{0.0}(f-n)=0.6$ and IRX$\sim 0.7$ where many of the galaxies are located. However, it is clear that the galaxies in our sample, while still obeying a tight relation between IRX and $^{0.0}(f-n)$, have a somewhat steeper slope than has been proposed in the past.  In particular, in agreement with \citet{gil_de_paz}, the galaxies with very low IRX, usually dwarf galaxies of low luminosity, tend to fall well below the typical relations. However, several complications, in addition to the calibration issues, must be mentioned when comparing this sample with others:

\begin{enumerate}

\item Bolometric corrections. --  We omit the IR bolometric correction ($BC_{IR}=1.75$) used by \citet{seibert05a} and \citet{MHC} since we have calculated total dust luminosities using the IR template SEDs of \citet{dale01} instead of the definition of \citet{helou88}. Similarly, \citet{kong04}, using a formula derived from the \citet{dale02} template SEDs, report IR luminosities $\sim 50\%$ higher than obtained using the prescription of \citet{helou88}, and they find a steeper IRX-$\beta$ relation than \citet{MHC} for the same sample of galaxies.  Moreover, as mentioned in \citet{paper1} significant variations in the $L(24\micron)$ to $L_{dust}$ ratios are possible in the models of \citet{dale01}, and these add to the systematic uncertainty in IRX, especially given that the mid- to far-IR colors of the sample galaxies do not closely match the templates used for the IR bolometric correction \citep[Fig. 1 of][]{paper1}.  Finally, the IRX that we use includes any contribution to IR from dust heated by old stars -- in particular light absorbed in the optical as opposed to the UV \citep{paper2} -- which can cause IRX to overestimate the UV attenuation.

\item $K$-corrections.-- A further complication is that the \citet{MHC} and \citet{seibert05a} relations are defined at $z=0$ and we must account for the effect of the $K$-correction to $z=0.1$ on the $f-n$ colors, and to a lesser extent on IRX (due to the dependence of IRX on $f$. The extrapolations required to go from $z=0.1$ (the median redshift of our sample) to $z=0.0$ can be nearly as large as the size of the scatter in the relation: $\langle ^{0.0}(f-n)-^{0.1}(f-n)\rangle =0.13$ mag with a $1\sigma$ dispersion of 0.17 mag, and individual $K$-corrections from the observed frame $f-n$ color to $^{0.0}(f-n)$ can be larger than 0.5 mag.  Note also that the method of \citet{blanton_k} relies on spectral synthesis modeling of the very type that we are testing. \citet{burgarella_g1} have suggested, on the basis of \emph{GALEX} grism spectra and photometry, that the observed frame $f-n$ color is a poor tracer of $\beta$ for galaxies with $z\gsim0.1$ - we find no significant difference in the distribution of galaxies in Figure \ref{fig:irx_beta_z0} if we further restrict the sample to $z<0.1$ 

\end{enumerate}

Note also that the \citet{MHC} and \citet{kong04} relations may have been affected by aperture effects in the IUE data (M Seibert 2007, private communication), and that the \emph{GALEX} color zero-point error is $\sim 0.1$ mag.  These circumstances highlight the difficulty in using the IRX-$\beta$ relation at high redshift, since it is poorly defined even at low redshift and subject to numerous systematic effects, as well as random measurement errors that are very important given the steepness of the relation.

In Figure \ref{fig:irx_beta_delts} we show the offset (in the IRX direction) of the sample galaxies from the \citet{kong04} relation (eq. \ref{eqn:kong}) as a function of \dfn. We also show the offsets of \airx~from the relation of \citet{kong04} for A$_{fuv}$. There is a trend of these offsets with \dfn, as suggested by \citet{kong04} as a signature of the effect of SFH on the UV color, although the trend is weak or non-existent for \dfn$<1.6$. The dashed lines in Figure \ref{fig:irx_beta_delts} show the predictions of \citet{kong04} (Eq 8 and 9) for two extreme values of $^{0.0}(f-n)$.  While there is some agreement with these predictions at $1.4<$\dfn$<1.8$ (modulo a constant offset), the predictions do not simultaneously reproduce the offsets at very low \dfn. The interpretation is futher complicated by the poor correspondence of the \citet{kong04} relation to the galaxies in this sample, evidenced by a large offset in Figure \ref{fig:irx_beta_z0}.


\begin{figure*}{t}
\plotone{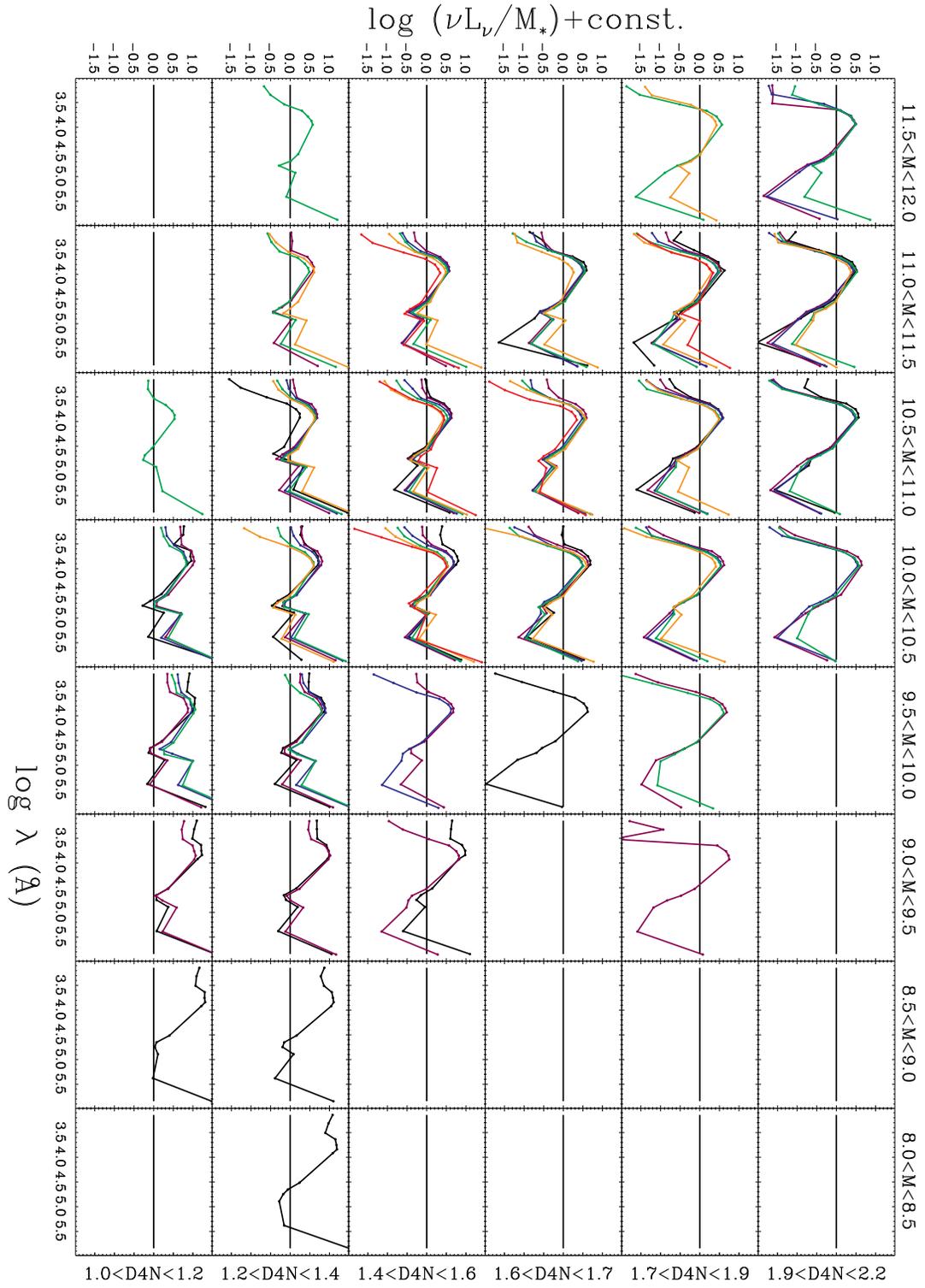}
\epsscale{1}
\figcaption{SEDs for different bins of stellar mass (given as $\log (\mbox{M}_{*}/\mbox{M}_{\odot})$) and \dfn, analogous to a color magnitude diagram.  Within each of these stellar mass and \dfn~bins the SEDs are normalized by stellar mass and further binned by \airx.  See text for details and the definition of the parameter bins. $\log (\mbox{M}_{*}/\mbox{M}_{\odot})$ increases from left to right, while \dfn~increases from bottom to top.  The color of each average SED encodes the \airx~bin ($black, purple, blue, green, orange$, and $red$, respectively) for the six bins defined by \airx$=$[0.0,1,2.0,2.5,3.5,4.5,5.5].
\label{fig:sedgrid}}
\end{figure*}

\section{Average Spectral Energy Distributions}
\label{sec:sed}
We have so far considered only the variation in the UV color with IRX. \citet{paper1} consider additional colors and their variation with IRX. However, it is possible to construct UV through IR spectral energy distributions (SEDs) for the sample galaxies that simultaneously show all colors, including the UV and IR behavior used to infer the SFR and attenuation.  We construct average SEDs for the sample galaxies by first normalizing the galaxy flux at all wavelengths by M$_*$ as given by \citet{kauffmann03a}. We then split the galaxies into bins of different galaxy parameters: IRX, \dfn, and M$_*$.  These SEDs are shown in Figure \ref{fig:sed}, and are given in Table \ref{table:sed}. Note that we have required a 70\micron~observation (i.e. only Lockman Hole galaxies are included) but we do not impose a S/N cut at this wavelength; the 70\micron~points in the SEDs, especially for red-sequence galaxies, should thus be treated with significant caution. These SEDs will be useful for comparison to high redshift galaxies and for theoretical studies. 

All three parameters -- IRX, \dfn, and M$_*$ -- appear correlated with significant change of the SED shapes. The scatter in SED shape is lowest when the SEDs are binned by \dfn.  The change of SED shape as a function of M$_*$ is driven by the change in SED shape with IRX and \dfn -- M$_*$ is correlated with both of these parameters. IRX and \dfn~are only slightly correlated with each other in the galaxies in this sample \citep[Fig. 7 of][]{paper1}, so the change in SED shapes with IRX and \dfn~indicate that both attenuation and SFH serve to influence the SED shape. This is shown for individual colors (instead of the whole SED) in \citet{paper1}.  Note that there is a dependence of the 3.6\micron~stellar mass to light ratio on \dfn, but little dependence of that ratio on IRX.

To show the separate effects of  IRX, \dfn, and M$_*$ on the SED shapes we can bin the large sample of galaxies by all three parameters simultaneously (after normalizing by M$_*$).  The results of this process are shown in Figure \ref{fig:sedgrid} where instead of using quintiles to define the parameter bins we set definite bins of width 0.5 in $\log$M$_*$ and for \dfn we use bin edges \dfn$=$[1.0,1.2,1.4,1.6,1.7,1.9,2.2]. These bins are arranged to approximate the color-magnitude diagram \citep[Fig. 2 of][]{paper1}. Within each of these stellar mass and \dfn~bins we further separate the galaxy SEDs into bins of \airx~with bin edges \airx$=$[0.0,1,2.0,2.5,3.5,4.5,5.5] (shown in Fig. \ref{fig:sedgrid} as $black, purple, blue, green, orange$, and $red$, respectively). Many of the SEDs shown are still averages of several galaxies, although at the low stellar mass end there may be only one galaxy per bin of stellar mass, \dfn, and \airx.  It is clear that for a given narrow range of \dfn~and M$_*$ there is a significant change in the SED shape due to varying \airx.  At higher stellar mass there is a larger number of significantly attenuated galaxies, and of course galaxies with higher \dfn~also tend to have more stellar mass, accounting for much of the change in the SED shape as a function of M$_*$ (but see \S\ref{sec:dust_law}).

\begin{figure}
\epsscale{1.0}
\plotone{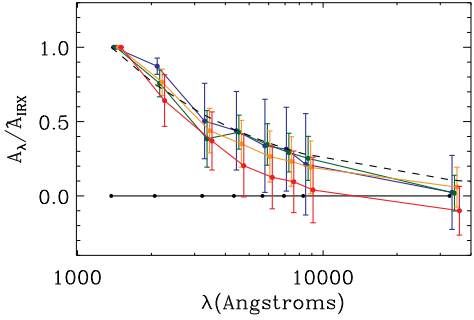}
\figcaption{Derived A$_{\lambda}/$\airx~(i.e. the dust attenuation law, \S\ref{sec:dust_law}) for different ranges of \dfn~(\dfn$=[1.1,1.2,1.3,1.4,1.5]$; $blue to red$).  The dashed line shows a $\lambda^{-0.7}$ attenuation curve. Galaxies are restricted to $9.5 < \log (\mbox{M}_{*}/\mbox{M}_{\odot}) < 11.5$.  The curves are shifted slightly in wavelength for clarity.
\label{fig:dust_law}}
\end{figure}

\section{An Empirically Derived Dust Attenuation Law}
\label{sec:dust_law}
The change in average SEDs as a function of IRX shows the effects of different amounts of attenuation on the stellar mass to light ratio, which increase for shorter wavelength.  The form of this increase shows, in effect, the average attenuation law for galaxies. It is possible to derive this attenuation law from these data by making several assumptions.  It is also necessary to simultaneously split the galaxies into narrow ranges of \dfn, since the SFH also has a strong effect on the color of galaxies.  We assume an optical depth:

\begin{equation}
\tau_{\lambda}=C \cdot G(\lambda)
\end{equation}

\noindent where we assume that $G(\lambda)$ is universal (the shape of the attenuation curve) for a given \dfn~and $C$ provides the normalization of the attenuation curve.  These are separable under the additional assumption of reasonably simple geometries. In principle this assumption can be tested by repeating the analysis detailed below for different ranges of \airx. Then the transmitted flux is


\begin{equation}
 f_\lambda=f_{o}e^{-\tau_{\lambda}}
\end{equation}

\noindent where $f_{o}$ is the intrinsic, unattenuated flux. So, 

\begin{eqnarray*}
A_{\lambda} & = & 2.5\log(f_{o}/f_\lambda) \\
 & = & 1.086\cdot C\cdot G(\lambda) 
\end{eqnarray*}

\noindent and the attenuation at a given wavelength may then be written in terms of the attenuation at another wavelength as $A_{\lambda_{2}}=A_{\lambda_{1}}(G(\lambda_{2})/G(\lambda_{1}))$.

The color of a galaxy, $c_{\lambda_{1},\lambda_{2}}=m_{\lambda_{1}}-m_{\lambda_{2}}$ where $m_{\lambda}$ is the magnitude at wavelength $\lambda$,  is given by the intrinsic color $c_o$ plus the difference in the effective attenuation at each wavelngth.

\begin{eqnarray*}
c_{\lambda_{1},\lambda_{2}} & = & c_o+(A_{\lambda_{1}}-A_{\lambda_{2}}) \\
 & = & c_o+A_{\lambda_{1}}[1-G(\lambda_{2})/G(\lambda_{1})]
\end{eqnarray*}


\begin{figure*}[t]
\plotone{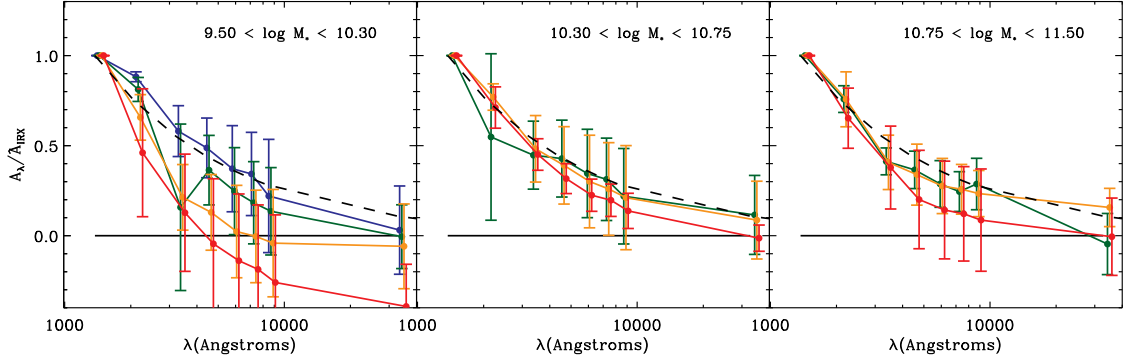}
\figcaption{As in Figure \ref{fig:dust_law}, the derived A$_{\lambda}/$\airx~(\S\ref{sec:dust_law}) for different ranges of \dfn~(\dfn$=[1.1,1.2,1.3,1.4,1.5]$); $blue to red$) and stellar mass.  Each panel shows a different stellar mass range.  We only show those stellar mass and \dfn~ranges for which there are $>10$ galaxies.  The dashed line shows a $\lambda^{-0.7}$ attenuation curve.   The curves are shifted slightly in wavelength for clarity.
\label{fig:dust_law_mass}}
\end{figure*}

We make the (large) assumption that \airx$=A_{\lambda_{1}}=A_{1390\AA}$. Assuming that $c_o$ is relatively constant between galaxies for narrow ranges of \dfn, the slope of the relation between the $^{0.1}(f-m_{\lambda_{2}})$ color and \airx, for a narrow range of \dfn, then gives $(1-G(\lambda_{2})/G(1390\AA))$.  This relation is effectively shown in \citet{paper1} for several different colors and ranges of \dfn, although IRX was used instead of \airx~and many of the colors used $^{0.1}n$ instead of $^{0.1}f$.  We split the sample into bins of \dfn(=[1.0,1.2,1.3,1.4,1.5]) and determine the slopes for each color using the ordinary least squares bisector \citep{isobe}. In Figure \ref{fig:dust_law} we show the resulting derived $G(\lambda_{2})/G(1390\AA)$ for the different ranges of \dfn.  Errors are determined from the range in the slopes derived by fitting \airx~given $^{0.1}(f-m_{\lambda_{2}})$ and vice-versa.  The resulting attenuation curves are consistent with a $\lambda^{-0.7}$ attenuation law. 

There are a number of caveats to this method for the determination of the dust attenuation law. Primarily, \airx~is not the true $A_{1390}$, especially for galaxies with redder intrinsic spectra. This is likely the reason for the differing shapes of the attenuation curves for different ranges of \dfn.  This analysis does not exploit the unique advantage of IRX as a dust indicator, i.e. that the IR emission constrains the total absorbed flux.  Indeed, the assumption that \airx$=A_{1390}$ is inconsistent with the fact that the derived attenuation laws are non-zero at longer wavelengths, which means that galaxies with different \dfn~are likely to have different $\eta$ (Eq \ref{eqn:airx} and \citet{paper2}). Second, $G(\lambda)$ may not be separable from $C$ (i.e. the shape of the attenuation curve may be dependent on the total amount of attenuation; this can happen due to geometric effects as shown by \citet{WG00}).  Another possible systematic error is that the unreddeded spectra may be determined by more than just \dfn~(e.g. metallicity, contamination due to aperture effects) -- $c_o$ may not be constant even in small ranges of \dfn, especially for colors based on the flux far from the 4000\AA~break.  

Another caveat to the method described above is that $G(\lambda)$ may not be universal, for example because of a variation in the extinction law between galaxies.  Such a variation might be expected if the composition of the dust causing the attenuation varies between galaxies.  Dust composition may vary with metallicity, which is correlated with stellar mass.  In particular the 2175\AA~bump in the Milky Way extinction curve, which is not seen in the LMC and SMC extinction curves, is thought to be due to PAH molecules \citep{draine_li06} -- these molecules have been shown to be underabundant in low-metallicity dwarf galaxies, although the reason for this is still unclear.  A variation in the attenuation law with mass would be of special importance for interpreting the location of galaxies in the specific SFR - stellar mass plane in terms of SFHs \citep{labbe07, noeske07a} and for comparison of UV to \ha~derived SFRs \citep{salim07}.  For these reasons we consider the attenuation laws derived as above but for bins of \dfn~\emph{and stellar mass} (i.e.  cells in Figure \ref{fig:sedgrid}).  These are shown in Figure \ref{fig:dust_law_mass}.  The smaller number of galaxies in each bin results in a weaker constraint on $G(\lambda_{2})/G(1390\AA))$ than in Figure \ref{fig:dust_law}. No significant trends with stellar mass are seen, although the uncertainties are large. A larger sample of galaxies is required to accurately test this hypothesis and, in addition, split the sample into, e.g., bins of \airx~to explore the coupling of attenuation law shape to the amount of attenuation in galaxies.

\section{Attenuation as a Function of Metallicity and Surface Brightness}
\label{sec:sigma_z}

At the local level the amount of dust attenuation is due to the column density of dust, $\tau=\int\sigma_{dust}n_{dust}\ell$ where $\sigma_{dust}$ is the absorption cross section, $n_{dust}$ is the dust volume density, and $\ell$ is the unit path length along the line of sight \citep[e.g.][]{WH96}.  Simplistically, one might write $n_{dust}\ell=\Sigma_{gas} f_{dust}$ where $f_{dust}$ is the global dust to gas ratio and $\Sigma_{gas}$ is the column density of gas along the line of sight to the star or cluster of stars. We assume that $f_{dust}\propto Z$, where $Z$ is the gas-phase metallicity. We obtain $Z$ from the $\log$O/H values of \citet{tremonti04} via the solar values $(\log$O/H$)_{\odot}=8.69$ and $Z_{\odot}=0.02$, assuming that the oxygen abundance is a tracer of the total metal abundance within a factor of 2 or 3.  In semi-analytic modelling it is typical to make the assumption $f_{dust}\propto Z$, and furthermore to use the global values of gas density and metallicity \citep{somerville01, martin07}.  This approach involves several other assumptions: that the effect of stars being embedded \emph{within} the gas can be neglected, that spatial variations in both gas surface density and metallicity are such that their product averages to the global value, and that all stars (of all ages) are affected equally.  Such a relation between surface density, metallicity, and attenuation has been investigated empirically, using global quantities of resolved galaxies, by \citet{boissier06} and \citet{cortese06}.  

\begin{figure*}[t]
\plotone{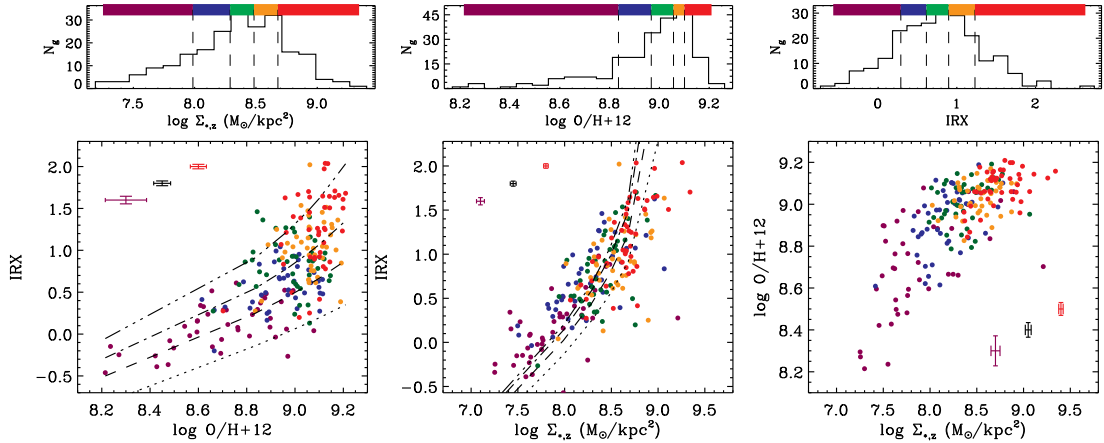}
\figcaption{Three projections of the relation between IRX, gas-phase metallicity, and surface mass density for the 219 galaxies of the sample for which metallicity measurements are available from \citet{tremonti04}.  In each large panel the color coding is given by the quintiles of the distribution shown above that panel.  In the left panel the black lines show ($bottom to top$) the relation of IRX to metallicity expected from eq. \ref{eqn:bell} for the values of $\Sigma_{*,r}$ that define, in increasing order, the limits of the bins used for the color coding.  In the middle panel the lines are similar but here the metallicity is fixed at the limits of the bins used for the color coding.
\label{fig:sigma_z}}
\end{figure*}

We consider here the relation of the globally measured attenuation to the global metallicity and surface density for the sample galaxies.  We do not have measurements of the gas surface density, and use instead the stellar mass surface density.  Following \citet{kauffmann03c} this is calculated as $\Sigma_{*,z}=0.5M_{*}/(\pi R_{50}^2)$ where $M_{*}$ is given by \citet{kauffmann03a} and $R_{50}$ is the radius (in kpc) enclosing 50\% of the Petrosian light in the $z$ band. \citet{bell03_radio}, Appendix B, \citep[see also][Appendix A.1]{calzetti07} makes a quantitative estimate of the relation in this case, obtaining:
\begin{equation}
\label{eqn:bell}
\tau_{fuv,1550}=1.7\eta Zf_{g}/(1-f_{g})\Sigma_{*}
\end{equation}
where $\tau_{fuv,1550}$ is the optical depth at 1550\AA, $Z$ is the metallicity, $f_{g}$ is the gas fraction, $\Sigma_{*}$ is the stellar mass surface density, and $\eta$ is a factor of order unity (\citet{bell03_radio} determine $\eta=0.7$ to fit the attenuation-luminosity correlation) to account for, e.g., relative star dust geometry.  Using the approximation $\tau_{fuv}=\tau_{fuv,1550}(1390/1550)^{-0.7}$ and equation 1, $\tau_{fuv,1550}$ can be related to IRX.

In Figure \ref{fig:sigma_z} we show three projections of the relation between IRX, gas-phase metallicity, and $\Sigma_{*,z}$. The number of galaxies for which this is possible is small (219), due to the lack of metallicity measurements -- these galaxies have strong emission lines and typically have large ratios of recent to past star formation. Note also that AGN identified in the optical spectra on the basis of emission line ratios are excluded from this sample with metallicity measurements \citep{tremonti04}. The left and center panels show that there is a strong correlation between IRX and $\Sigma_{*,z}$, even when the metallicity is held nearly constant; metallicity and surface mass density are also correlated with each other, as shown in the right panel.  \citet{cortese06} found a similar trend of IRX with H-band surface brightness (a close proxy for stellar mass surface density), although they did not simultaneously control for metallicity.  The correlation of IRX with metallicity is very weak, for constant surface mass density.  However, there is an overall correlation of IRX with metallicity induced primarily by the correlation of metallicity with surface mass density.  The lines in Figure \ref{fig:sigma_z} show those expected from equation \ref{eqn:bell}, for the quintiles of the parameter used for the color coding (i.e., points of a certain color should fall between two adjacent lines). We have assumed $f_{g}=0.1$ and require $\eta=0.7$ to fit the data reasonably -- this relation is very sensitive to the choice of $\eta$. The lines in the middle panel, for different quintiles of the metallicity distribution, show that the effect of the metallicity variations is \emph{expected} to be weaker than the surface mass density variations, since the range of metallicity is only 1 order of magnitude while the range in surface mass density is 2 orders of magnitude.  Note that the shallower-than-expected slope of the data in this panel may well be due to the anti-correlation of $f_{g}$, which we have assumed to be constant, with surface density.

Indeed, it is possible to estimate $\Sigma_{g}$ from the star formation rate density, assuming the relation of \citet{kennicutt_schmidt} holds globally for these galaxies.  Gas densities are calculated via 
\begin{equation}
\label{eqn:kennicutt}
\Sigma_{g,k}=10^6[(10^4/5)\mbox{SFR}_{e}/(\pi R_{50}^2)]^{-1.4} \mbox{M}_{\sun}\cdot kpc^{-2}
\end{equation}
using the median SFR derived by \citet{brinchmann04}, and $R_{50}$ as above (the conclusions are unchanged if we instead use the $u$ band half light radius as an estimate of the extent of star formation).  The mean of the ratio $\Sigma_{g,k}/\Sigma_{*,z}$ is approximately 0.16 (hence the assumption $f_{g}=0.1$ made above).  There is a significant trend of this ratio with surface mass density $\Sigma_{*,z}$, such that galaxies with larger $\Sigma_{*,z}$ have lower $\Sigma_{g,k}/\Sigma_{*,z}$ -- $\Sigma_{g,k}$ spans a much smaller range than  $\Sigma_{*,z}$.  This explains the steeper slope predicted by equation \ref{eqn:bell}.

These results may be compared to those of \citet{boissier06}, who found a correlation of IRX with metallicity \citep[see also][]{boissier04, heckman98}.  These studies did not consider the simultaneous effect of the surface density-metallicity correlation, and a trend with metallicity may thus be due to this additional component of the optical depth.  What is perhaps more puzzling is that \citet{boissier06} did not find a good correlation between the local IRX and the local HI(+H$_2$) surface density (but see \citet{buat96} and \citet{xu97}).  The relation between gas surface density and stellar surface density is highly scattered and poorly constrained, although we have assumed that the global values are simply related by $f_g$.  If the stellar surface mass density is better correlated with IRX than the gas surface density we must ask why. Is the stellar mass density tracing the \emph{history} of metal and dust production, such that higher densities naturally lead to higher attenuations?  This would seem to be accounted for by considering narrow ranges of metallicity.  One component of the analysis that we have treated only superficially is that stars are necessary to measure IRX and hence the attenuation -- a region of large gas and dust column will not contribute to the measured attenuation if there are few stars, and conversely a large number of stars in a low dust column region will cause a low measured IRX, even if the global gas and dust surface density is large.  IRX in fact measures the sum of the local $\Sigma_{gas}Z$ weighted by the local stellar luminosity.

This relation between surface-density, metallicity, and IRX is very closely related to the relations between IRX and luminosity and between IRX and stellar mass -- these, in turn, are key to interpreting the color-magnitude diagram of galaxies in terms of models for the build-up of stellar mass \citep{labbe07, noeske07a}.  If the sizes of late-type galaxies are nearly constant or correlated with the total SFR, then the gas surface densities are also correlated with total SFR.  In this case, the relation between IRX and gas surface density,  a more direct probe of the physical conditions related to attenuation, is almost trivially related to the well-known correlation between IRX and SFR \citep{martin05}.  A similar argument may be made for the stellar mass surface density and total stellar mass. Interestingly, the scatter in the relation presented here does not appear to be significantly less than the scatter in the IRX-L$_{SF}$ relation \citep[see][]{paper2}. 

If, however, the measured attenuation does prove a useful probe of the global $\Sigma_{gas}Z$, then the measured distribution of attenuation may provide a strong constraint on models of galaxy evolution, since both gas surface density and metallicity are intimately related to star birth and star death \citep{martin07}.

\section{Conclusions}

\begin{enumerate}

\item For \dfn~$< 1.7$ we find little evidence that scatter in the IRX-$\beta$ diagram is caused primarily by variations in UV color with the SFH of the stellar population, as suggested by \citet{kong04}.  However,  IRX may be affected by the SFH, which could mask such an effect. 

\item We assess the utility of the IRX-$\beta$ relation for determining dust attenuation at high redshift using UV colors alone.  While the scatter is small, the large slope makes the derived IRX very sensitive to even small errors in the UV color ($\sim 0.1$ mag).  These may be random (due to measurement errors) or systematic, for example due to the method used to $K$-correct the data. Furthermore, significant uncertainty in the determination of the bolometric IR luminosity makes even the low redshift relation uncertain.

\item We use the UV through IR data to show how the entire spectral energy distribution (not just a single color) varies as a function of the relevant galaxy parameters IRX, \dfn, and M$_{*}$.  This results in high-quality, low resolution 1375\AA~through 70\micron~average SEDs normalized by M$_{*}$ for different ranges of IRX, \dfn, and M$_{*}$.  Such average SEDs show a variation of the 3.6\micron~mass to light ratio with \dfn, when stellar masses are estimated from optical data following \citet{kauffmann03a}.  They also show the strong variation in the UV and IR specific luminosities with these quantities, although the variation with M$_{*}$ is driven largely by the correlation of SFH and attenuation with M$_{*}$ (which might be expected to change with redshift). 

\item We have used these average SEDs to derive a low resolution dust attenuation curve for blue-sequence galaxies, for several different ranges of \dfn.  The accuracy of the attenuation curves is limited by the variation of the relation between IRX and the true attenuation $A_{fuv}$, but the derived attenuation curves are consistent with a $\lambda^{-0.7}$ law.  To investigate possible changes in the attenuation law as a function of metallicity or PAH abundance we repeat the analysis, splitting galaxies by their stellar mass.  We find no significant differences as a function of mass although the uncertainties are large. 

\item  IRX is correlated with both stellar mass surface density and gas-phase metallicity (as determined by \citet{kauffmann03c} and \citet{tremonti04}, respectively), although the latter correlation appears weak for a given narrow range of stellar mass surface density.  To the extent that the stellar mass surface density is related to the gas surface density, IRX then probes the causes of dust attenuation and this relation will allow the accurate calculation of attenuation in models of galaxy formation. This, in turn, will allow for the consistent comparison of photometric data to these models. Conversely, the attenuation can be used to probe the variation in the surface density and metallicity as a function of time through observations of high-redshift galaxies in the restframe UV and Mid to Far-IR.

\end{enumerate}

\acknowledgments
BDJ would like to thank A. Basu-Zych, S. Salim, A. Boselli, S. Boissier, and L. Cortese for comments that improved the paper. BDJ was supported by NASA GSRP Grant NNG-05GO43H

\emph{GALEX} (Galaxy Evolution Explorer) is a NASA Small Explorer, launched in April 2003. We gratefully acknowledge NASA's support for construction, operation, and science analysis for the \emph{GALEX} mission, developed in cooperation with the Centre National d'Etudes Spatiales of France and the Korean Ministry of Science and Technology. 

This work is based in part on observations made with the Spitzer Space Telescope, which is operated by the Jet Propulsion Laboratory, California Institute of Technology under a contract with NASA.  In particular, the publicly available \emph{Spitzer} data obtained by the SWIRE team have been essential to this work.

Funding for the SDSS and SDSS-II has been provided by the Alfred P. Sloan Foundation, the Participating Institutions, the National Science Foundation, the U.S. Department of Energy, the National Aeronautics and Space Administration, the Japanese Monbukagakusho, the Max Planck Society, and the Higher Education Funding Council for England. The SDSS Web Site is http://www.sdss.org/. The SDSS is managed by the Astrophysical Research Consortium for the Participating Institutions. The Participating Institutions are the American Museum of Natural History, Astrophysical Institute Potsdam, University of Basel, University of Cambridge, Case Western Reserve University, University of Chicago, Drexel University, Fermilab, the Institute for Advanced Study, the Japan Participation Group, Johns Hopkins University, the Joint Institute for Nuclear Astrophysics, the Kavli Institute for Particle Astrophysics and Cosmology, the Korean Scientist Group, the Chinese Academy of Sciences (LAMOST), Los Alamos National Laboratory, the Max-Planck-Institute for Astronomy (MPIA), the Max-Planck-Institute for Astrophysics (MPA), New Mexico State University, Ohio State University, University of Pittsburgh, University of Portsmouth, Princeton University, the United States Naval Observatory, and the University of Washington.



{\it Facilities:} 

\bibliographystyle{apj}
\bibliography{ms}



\clearpage

\begin{landscape}

\begin{deluxetable}{cc|ccccccccccccc}
\tabletypesize{\tiny}
\setlength{\tabcolsep}{1mm}
\tablecolumns{15}
\tablewidth{0pc}
\tablecaption{Average\tablenotemark{a} Broadband SEDs Normalized by M$_*$ in Bins of IRX, \dfn, and M$_*$ \label{table:sed}
}
\tablehead{}
\startdata

\multicolumn{2}{c}{Bin} & \multicolumn{13}{c}{ $<\log (\nu F_{\nu}/M_{*})+C\tablenotemark{b}> (\sigma[\log (\nu F_{\nu}/M_{*})])$}\\
 Min. & Max. & $^{0.1}f$ & $^{0.1}n$ & $^{0.1}u$ & $^{0.1}g$ & $^{0.1}r$ & $^{0.1}i$ & $^{0.1}z$ & $^{0.1}3.6\micron$ &  $4.5\micron$ & $5.8\micron$ & $7.8\micron$ & $24\micron$ & $70 \micron$ \\
\cutinhead{Binned by IRX}
-0.46 &  0.61 &  0.19 (0.467) &  0.18 (0.395) &  0.30 (0.201) &  0.66 (0.115) &  0.77 (0.071) &  0.81 (0.061) &  0.78 (0.042) &  0.06 (0.035) & -0.22 (0.035) & -0.32 (0.074) & -0.03 (0.197) & -0.59 (0.369) &  0.76 (0.415) \\
 0.61 &  0.87 & -0.44 (0.456) & -0.39 (0.382) & -0.09 (0.264) &  0.39 (0.062) &  0.56 (0.034) &  0.62 (0.029) &  0.64 (0.021) & -0.03 (0.027) & -0.28 (0.030) & -0.46 (0.058) & -0.21 (0.272) & -0.80 (0.404) &  0.46 (0.468) \\
 0.87 &  1.09 & -0.71 (0.551) & -0.57 (0.424) & -0.11 (0.113) &  0.35 (0.050) &  0.54 (0.029) &  0.60 (0.024) &  0.62 (0.019) & -0.02 (0.026) & -0.27 (0.033) & -0.44 (0.069) & -0.23 (0.349) & -0.83 (0.514) &  0.49 (0.549) \\
 1.09 &  1.39 & -0.95 (0.414) & -0.76 (0.307) & -0.23 (0.087) &  0.27 (0.043) &  0.48 (0.026) &  0.55 (0.023) &  0.58 (0.017) & -0.04 (0.021) & -0.28 (0.027) & -0.48 (0.053) & -0.30 (0.289) & -0.79 (0.376) &  0.52 (0.529) \\
 1.39 &  2.63 & -1.17 (0.300) & -0.89 (0.194) & -0.32 (0.061) &  0.19 (0.033) &  0.41 (0.023) &  0.49 (0.021) &  0.52 (0.017) & -0.03 (0.024) & -0.27 (0.036) & -0.42 (0.058) & -0.13 (0.242) & -0.54 (0.292) &  0.77 (0.366) \\
\cutinhead{Binned by \dfn}
 1.06 &  1.31 &  0.37 (0.201) &  0.39 (0.158) &  0.46 (0.095) &  0.77 (0.068) &  0.84 (0.048) &  0.88 (0.042) &  0.85 (0.030) &  0.20 (0.020) & -0.06 (0.022) & -0.09 (0.032) &  0.43 (0.056) &  0.01 (0.080) &  1.38 (0.083) \\
 1.31 &  1.45 & -0.18 (0.206) & -0.12 (0.139) &  0.08 (0.066) &  0.45 (0.040) &  0.59 (0.027) &  0.65 (0.022) &  0.65 (0.017) &  0.03 (0.013) & -0.21 (0.017) & -0.31 (0.021) &  0.18 (0.041) & -0.29 (0.080) &  0.97 (0.116) \\
 1.45 &  1.66 & -0.63 (0.262) & -0.50 (0.151) & -0.16 (0.063) &  0.30 (0.034) &  0.49 (0.025) &  0.56 (0.020) &  0.58 (0.016) & -0.05 (0.010) & -0.29 (0.016) & -0.46 (0.020) & -0.12 (0.054) & -0.62 (0.094) &  0.69 (0.132) \\
 1.66 &  1.85 & -1.09 (0.291) & -0.89 (0.158) & -0.35 (0.071) &  0.20 (0.020) &  0.43 (0.016) &  0.51 (0.013) &  0.55 (0.011) & -0.10 (0.010) & -0.35 (0.015) & -0.59 (0.015) & -0.52 (0.085) & -1.10 (0.140) &  0.17 (0.294) \\
 1.85 &  2.20 & -1.52 (0.118) & -1.28 (0.078) & -0.47 (0.103) &  0.13 (0.015) &  0.39 (0.012) &  0.47 (0.011) &  0.52 (0.009) & -0.14 (0.014) & -0.40 (0.020) & -0.67 (0.019) & -0.86 (0.054) & -1.51 (0.120) & -0.17 (0.255) \\
\cutinhead{Binned $\log$M$_*$}
 0.00 &  9.70 &  0.54 (0.557) &  0.55 (0.356) &  0.59 (0.317) &  0.93 (0.087) &  0.99 (0.054) &  1.02 (0.045) &  0.97 (0.032) &  0.20 (0.033) & -0.11 (0.041) & -0.10 (0.070) &  0.13 (0.165) & -0.29 (0.277) &  1.13 (0.280) \\
 9.70 & 10.50 & -0.29 (0.667) & -0.16 (0.416) &  0.14 (0.126) &  0.53 (0.069) &  0.67 (0.039) &  0.72 (0.032) &  0.71 (0.024) &  0.06 (0.026) & -0.22 (0.033) & -0.30 (0.060) &  0.06 (0.257) & -0.43 (0.325) &  0.86 (0.488) \\
10.50 & 10.75 & -0.58 (0.446) & -0.46 (0.309) & -0.10 (0.086) &  0.34 (0.040) &  0.53 (0.021) &  0.59 (0.017) &  0.61 (0.013) & -0.01 (0.016) & -0.27 (0.024) & -0.43 (0.045) & -0.11 (0.237) & -0.66 (0.329) &  0.65 (0.406) \\
10.75 & 11.00 & -0.84 (0.385) & -0.69 (0.284) & -0.22 (0.068) &  0.27 (0.030) &  0.47 (0.019) &  0.54 (0.016) &  0.57 (0.013) & -0.05 (0.019) & -0.28 (0.029) & -0.49 (0.040) & -0.28 (0.244) & -0.79 (0.365) &  0.50 (0.415) \\
11.00 & 12.50 & -1.12 (0.300) & -0.97 (0.220) & -0.40 (0.122) &  0.15 (0.023) &  0.39 (0.015) &  0.46 (0.013) &  0.50 (0.010) & -0.13 (0.013) & -0.33 (0.026) & -0.59 (0.033) & -0.49 (0.194) & -1.08 (0.347) &  0.23 (0.366) \\
\enddata
\tablenotetext{a}{Because $\nu F_{\nu}/$M$_{*}$ is close to log-normally distributed for the parameter bins given here, the mean of the log of this quantity is an appropriate statistic to report, as opposed to the mean or the log of the mean.}
\tablenotetext{b}{$C=0.99$}
\end{deluxetable}
\clearpage
\end{landscape}






\end{document}